\documentclass[a4paper,11pt]{article}
\usepackage{pos}
\usepackage{booktabs}

\title{Quark and gluon momentum fractions in the pion and in the kaon}
\author[a,b]{Constantia Alexandrou}
\author[a]{Simone Bacchio}
\author[c]{Martha Constantinou}
\author[c]{Joseph Delmar}
\author[d]{Jacob Finkenrath}
\author[e]{Bartosz Kostrzewa}
\author[e,f]{Marcus Petschlies}
\author*[a,g]{Luis Alberto Rodriguez Chacon}
\author[b]{Gregoris Spanoudes}
\author[e,f]{Fernanda Steffens}
\author[e,f]{Carsten Urbach}
\author[h]{Urs Wenger}

\affiliation[a]{Computation-based
Science and Technology Research Center, The Cyprus Institute, 20 Kavafi Str., Nicosia 2121, Cyprus}

\affiliation[b]{Department of Physics, University of Cyprus, P.O. Box 20537, 1678 Nicosia, Cyprus}

\affiliation[c]{Department of Physics,  Temple University,  Philadelphia,  PA 19122 - 1801,  USA}

\affiliation[d]{Department of Theoretical Physics, European Organization for Nuclear Research, CERN,\\
1211, Genève, Switzerland}

\affiliation[e]{Helmholtz-Institut für Strahlen- und Kernphysik, University of Bonn, Germany}

\affiliation[f]{Bethe Center for Theoretical Physics, University of Bonn, Germany}

\affiliation[g]{Department of Physics and Earth Science, INFN, University of Ferrara, Via Saragat 1, 4122 Ferrara, Italy}

\affiliation[h]{Institute for Theoretical Physics, Albert Einstein Center for Fundamental Physics, University of Bern, Switzerland}

\emailAdd{l.a.chacon@cyi.ac.cy}
\abstract{ We present results on  the momentum fraction carried by quarks and gluons in the  pion and the kaon.  We employ three gauge ensembles generated  with $N_f=2+1+1$ Wilson twisted-mass clover-improved fermions with physical quark masses. We perform, for the first time, a continuum extrapolation directly at the physical pion. We find that the  total momentum fraction carried by quarks is $\langle x \rangle_{q, R}^{\pi}= 0.575(79)$ and $\langle x \rangle_{q,R}^{K} = 0.683(50)$ and by gluons $\langle x \rangle_{g, R}^{\pi}=0.402(53)$ and $\langle x \rangle_{g, R}^{K}=0.422(67)$ in the pion and in the kaon, respectively, in the $\MSbar$ scheme and at the renormalization scale of 2 GeV.
  Having computed both the quark and gluon contributions in the continuum limit,  we verify the momentum sum, finding  0.984(89) for the pion and 1.13(11) for the kaon.}

\FullConference{The 41st International Symposium on Lattice Field Theory (LATTICE2024)\\
 28 July - 3 August 2024\\
Liverpool, UK\\}

\newcommand{\pvec}{\mathbf{p}}
\newcommand{\qvec}{\mathbf{q}}
\newcommand{\MSbar}{\overline{\mathrm{MS}}}
\newcommand{\tins}{t_{\mathrm{ins}}}
\newcommand{\tsnk}{t_{\mathrm{s}}}

\newcommand{\avgx}{\langle x \rangle}

\begin{document}
\maketitle

\section{Introduction}

There is a great interest in understanding the inner structure of hadrons both experimentally and theoretically.  For instance, in the case of the proton, after the measurement of the quark intrinsic spin  by the European Muon Collaboration, which found  only around half of the total spin~\cite{EuropeanMuon:1989yki}, there has been a effort to understand the missing spin. 
Recent lattice QCD calculations showed that for the case of the proton not only the valence quarks contribute to the spin, but also the sea quarks and gluons~\cite{Alexandrou:2020sml}. 
The pion and kaon  being spin zero have a simpler structure  but they are  experimentally challenging. Experimental data is limited to studies from decades ago~\cite{Conway:1989fsj, Badier:1980jqm} which are still being used in global analysis for both the pion~\cite{Barry:2018ort,Novikov:2020xft,Barry:2021qpu,Barry:2022hph,Kotz:2023pdb}
 and kaon~\cite{Chen:2016pion,Shi:2018dyson,Lan:2020blfq,Bednar:2020dist,Cui:2020kaon,Han:2021maxent,Roberts:2021mass,Pasquini:2023pion}.  New experiments are being planned at  the future Electron-Ion Collider (EIC) and Electron-Ion Collider in China (EIcC), which  aim to measure quark and gluon contributions in the pion and kaon~\cite{Aguilar:2019teb, AbdulKhalek:2021gbh, Xie:2021ypc}. In this work, we present the first flavor decomposition of the average momentum fraction  using three gauge ensembles with $N_f = 2 + 1 + 1$ twisted-mass clover-improved fermions simulated with  physical quark masses and in the continuum limit.

% To calculate the momentum fraction of the pion and kaon  we need to evaluate the energy-momentum tensor
% \begin{equation}
%     \langle h(\textbf{p}) | T^{\mu\nu}_{X} | h(\textbf{p})\rangle = 2 \langle x \rangle_{X} \left( p_{\mu}p_{\nu} - \delta_{\mu\nu}\frac{p^2}{4} \right)
% \end{equation}

% {\bf Dina why do you the formula fo the nucleon??? Replace for spin zero particles.}
% \begin{equation}
% \begin{split}
%     \langle N(p', s') | T^{\mu\nu}_{X} | N(p, s)\rangle = \bar{u}_N(p',s') \Bigg[
%     A_{20}^{X}(q^2)\gamma^{\{\mu}P^{\nu\}} + 
%     B_{20}^{X}(q^2)\frac{i \sigma^{\{\mu\rho}q_{\rho}P^{\nu\}}}{2m_N} \\
%     + C_{20}^{X}(q^2)\frac{q^{\{\mu}q^{\nu\}}}{m_N}\Bigg] u_N(p,s),
% \end{split}
% \label{eq:EMT}
% \end{equation}

% where $X$ denotes the quark and gluon contributions. %{\bf Dina NO!!!!! do for pions and kaons, Luis. }

\section{Lattice Setup}

Three gauge field ensembles generated by the Extended Twisted Mass Collaboration (ETMC) with $N_f = 2 + 1 + 1$ quarks reproducing physical pion mass, are used in this work~\cite{Alexandrou:2018egz}. The parameters of these gauge ensembles are given in Table~\ref{tab:ensembles}. 
%{\bf Dina give reference}
\begin{table}[h]
  \centering
  \caption{ETMC ensembles analyzed. $a$ is the lattice spacing and $L\,(T=2L)$ the lattice spatial (temporal) extent in fm, and
    $M_{\pi^{\pm}}$ and $M_{K^{\pm}}$ the charged pion and kaon mass, respectively.}
\begin{tabular}{lccccc}
\hline
Ensemble & a [fm] & L~[fm] &  $M_{\pi^{\pm}}$~[MeV] & $M_{K^{\pm}}$~[MeV] \\
\hline
cB211.072.64 (B)&  $0.0796(1)$ & $5.09$  & $140.40(22)$  &  $498.41(11)$ \\
cC211.060.80 (C) &  $0.0682(1)$ & $5.46$  & $136.05(30)$   &  $495.27(14)$ \\
cD211.054.96 (D)&  $0.0569(1)$ & $5.46$  & $141.01(22)$  &  $494.77(11)$ \\
\hline
\end{tabular}
\label{tab:ensembles}
\end{table}

\noindent
The momentum fraction is extracted from the matrix element of the energy-momentum tensor in the forward direction given by
\begin{equation}
  \label{eq:x}
  \langle h(\pvec) \,|\, \bar{T}^X_{\mu\nu} \,|\, h(\pvec) \rangle =
  2\langle x\rangle_X\left(p_\mu p_\nu - \delta_{\mu\nu}\frac{p^2}{4}\right),
\end{equation}
where $h(\pvec)$ represents the pion or kaon state,  the index $X = q, g$ denotes the contributions from quarks and gluons to the total momentum of the hadron. The energy and momentum tensor  for quarks and gluons in Euclidean space-time is given, respectively, by:

\begin{align}
  \bar{T}_{\mu\nu}^q\ =\ -\frac{(i)^{\kappa_{\mu\nu}}}{4}\,\bar q\,
  \left(
    \gamma_\mu\stackrel{\leftrightarrow}{D}_\nu + \gamma_\nu \stackrel{\leftrightarrow}{D}_\mu
    - \delta_{\mu\nu}\,\frac{1}{2}\,\gamma_\rho\,\stackrel{\leftrightarrow}{D}_\rho
  \right)q\,,
  \nonumber \\
%%%
  \bar{T}^g_{\mu\nu} \ =\ (i)^{\kappa_{\mu\nu}}\,\left(
  F_{\mu\rho}^{~} F_{\nu\rho} +
  F_{\nu\rho}^{~} F_{\mu\rho} - \delta_{\mu\nu}\,\frac{1}{2}\,F_{\rho\sigma}{~}F_{\rho\sigma}
  \right) \,,
  \label{eq:Tdef}
\end{align}
with $\kappa_{\mu\nu} = \delta_{\mu,4}\,\delta_{\nu,4}$,
and the symmetrized covariant derivative
$ \stackrel{\leftrightarrow}{D_\mu} \, = \, \stackrel{\rightarrow}{D}_\mu - \stackrel{\leftarrow}{D}_\mu$ and 
 $F_{\mu\nu}$ is the gluon field-strength tensor.

\noindent
To compute the matrix elements in Eq.\ref{eq:x}, ratios of the following three- and two-point functions are analysed\vspace*{-0.6cm}

\begin{align}
  R_{44}^X(\tins, \tsnk) &= -\frac{4}{3}\,
  \frac{\omega_T(\tsnk,\mathbf{0})}{m_h} \, \frac{\langle h(\tsnk,\mathbf{0})\,\bar{T}^X_{44}(\tins)\,h(0,\mathbf{0})\rangle}
{\langle h(\tsnk,\mathbf{0})\ h(0,\mathbf{0})\rangle} \,,
  \nonumber \\
%%%
  R_{4k}^X(\tins, \tsnk) &=
  \frac{\omega_T(\tsnk,\pvec)}{\pvec^2} \, \frac{ \sum_{j}\,\pvec_j \, \langle h(\tsnk,\pvec)\,\bar{T}^X_{4j}(\tins)\,h(0,\pvec)\rangle}
  {\langle h(\tsnk,\pvec)\ h(0,\pvec)\rangle}\,,
  \label{eq:Rdef}
\end{align}
where $|\pvec| = 2\pi/L$, $\tsnk$ is the time separation between source and sink and $\tins$ the time separation between the source and the  operator insertion. The term $\omega_T(\tsnk,\qvec)$ in the numerator of Eq.~\ref{eq:Rdef}. If $\tsnk$ and $\tins$ are large enough, then
asymptotically
 both ratios of Eq.~(\ref{eq:Rdef}) are reduced
to $\langle x\rangle_X$.

\section{Analysis}

\begin{figure}[h!]
    \centering
    \includegraphics[width=\linewidth]{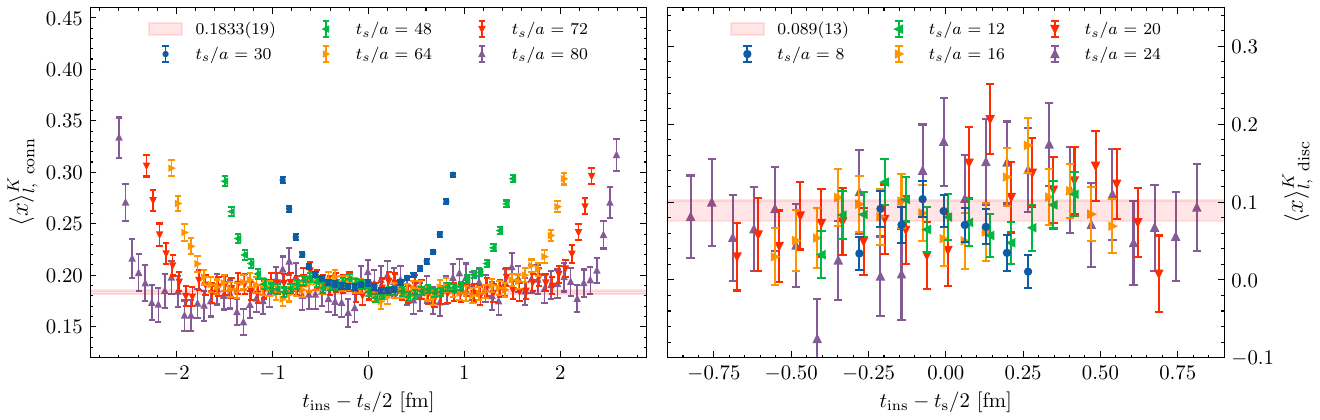}
    \caption{
      The bare ratio $R^{\ell}_{44}$ for quark-connected (left) and $R^{\ell}_{4k}$ for quark-disconnected (right)
      contributions to $\avgx^K_{\ell}$ versus the  $\tins-\tsnk/2$ for the C ensemble.
      In each case, we display results for several values of $\tsnk$, namely $\tsnk/a=30$ to $80$ for the quark-connected
      and $\tsnk/a=8$ to $24$
      for the quark-disconnected contribution. 
     The red band shows the model average value 
      from plateau fits  for varying combinations of $\tsnk$.}    
    \label{fig:ratio_quark}
\end{figure}

The ratios of Eq.~\ref{eq:Rdef} for  the connected and disconnected contributions are shown in Fig.~\ref{fig:ratio_quark}, where we show the bare 
ratio $R^{\ell}_{44}$ for the light  quark-connected as well as 
 $R^{\ell}_{4k}$ for light quark-disconnected. For the charged  kaon,  the light connected contribution is from the $u$-quark. The disconnected contribution shown in Fig.~\ref{fig:ratio_quark}, corresponds to the disconnected $u+d$ quark loop. The errors on the ratios are computed using a jackknife analysis.
Both quark-connected and disconnected ratios  converge to a constant for large values of the $\tins$ and $\tsnk$. Within this so-called  plateau region, we perform simultaneously fits to a constant using data sets for several combinations of $\tsnk$ and 
varying ranges of $\tins$. We model-average over these fits, with weights based on the Akaike Information Criterion~\cite{Jay:2020jkz}.

\begin{figure}
    \centering
    \includegraphics[width=\linewidth]{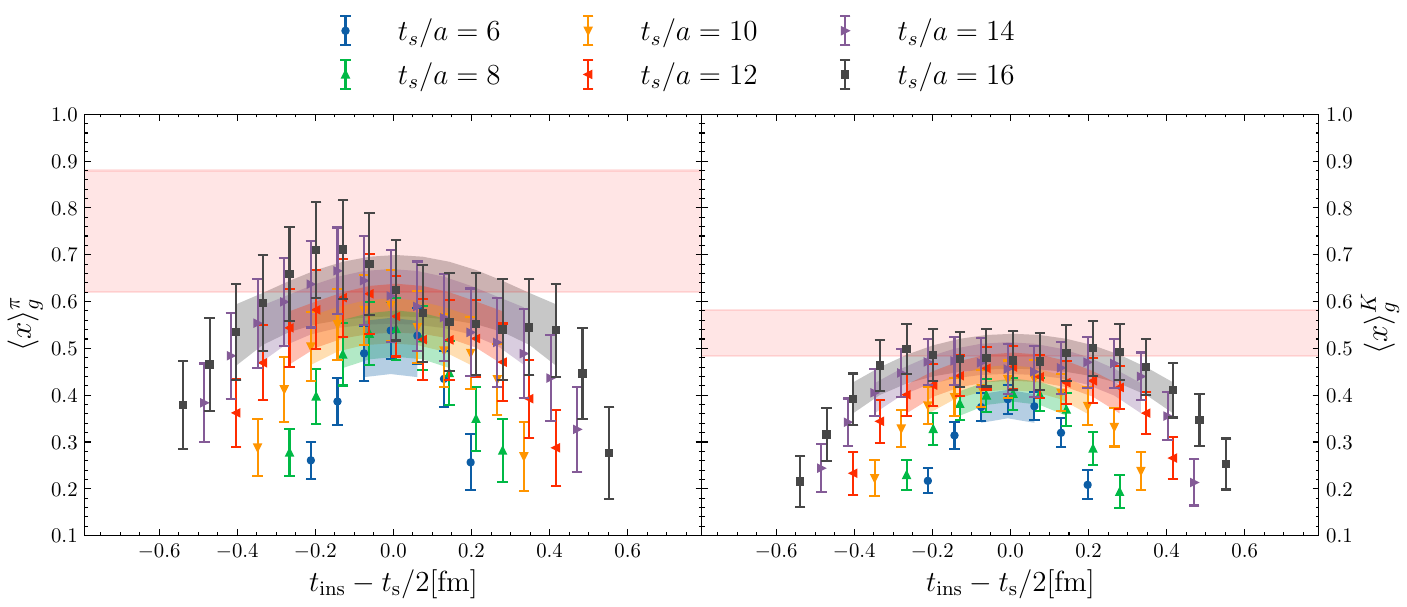}
    \caption{The bare ratios for the gluon contribution using the ensemble C. 
     We show data for six values of the sink-source  time separation $t_s$ 
     using 10 stout-smearing steps. The left panel is for the pion and the right panel for the kaon. The horizontal band in each plot shows the value after model-averaging over the constant and two-state fits and varying $\tsnk$ and $\tins$ included .}%{\bf Dina check}    
    \label{fig:ratio_gluon}
\end{figure}

In the case of the gluon contribution, we apply four-dimensional stout smearing \cite{Morningstar:2003gk} to the
gauge field used to construct the lattice gluon field-strength tensor. In Fig.~\ref{fig:ratio_gluon}, we show an example for the C ensemble for 10 stout smearing steps. As can be seen, there are excited state contributions and thus we perform both plateau and  two-state fits. After renormalization , the results are independent of the stout smearing in the range $5\le n_{stout}\le 10$, and therefore, the final result is obtained by fitting simultaneously ratios with $5\le n_{stout}\le 10$. We perform a model average of the results arising from varying the number of $\tsnk$ we include in the fit and from plateau and two-state fits.

\section{Renomalization and Continuum Limit}

The calculated quark and gluon average momentum fractions are renormalized nonperturbatively by using the RI$'$/MOM scheme followed by perturbative conversion to $\overline{\rm MS}$ at the reference scale of 2 GeV. The quark flavor-singlet and gluon components mix under renormalization according to:

\[
\begin{pmatrix}
\langle x \rangle_{q, \mathrm{R}} \\
\langle x \rangle_{g, \mathrm{R}}
\end{pmatrix}
=
\begin{pmatrix}
Z_{qq}^s & Z_{qg} \\
Z_{gq} & Z_{gg}
\end{pmatrix}
\begin{pmatrix}
\langle x \rangle_q \\
\langle x \rangle_g
\end{pmatrix}.
\]
The non-singlet combinations $\langle x \rangle_{u+d-2s}$ and $\langle x \rangle_{u+d+s-3c}$ are also calculated.  The renomalized quantities are given by:

\begin{align}
    \langle x \rangle_{u+d-2s} &= Z_{qq}(\langle x \rangle_u + \langle x \rangle_d - 2\langle x \rangle_s),\\
    \langle x \rangle_{u+d+s-3c} &= Z_{qq} (\langle x \rangle_u +\langle x \rangle_d + \langle x \rangle_s -3\langle x \rangle_c),
\end{align}
where $Z_{qq}$ denotes the nonperturbatively determined non-singlet renormalization factor. The calculation of all renormalization factors entering our study is described in detail in Ref.~\cite{alexandrou:pion_kaon}.

\begin{figure}
    \centering
    \includegraphics[width=0.8\linewidth]{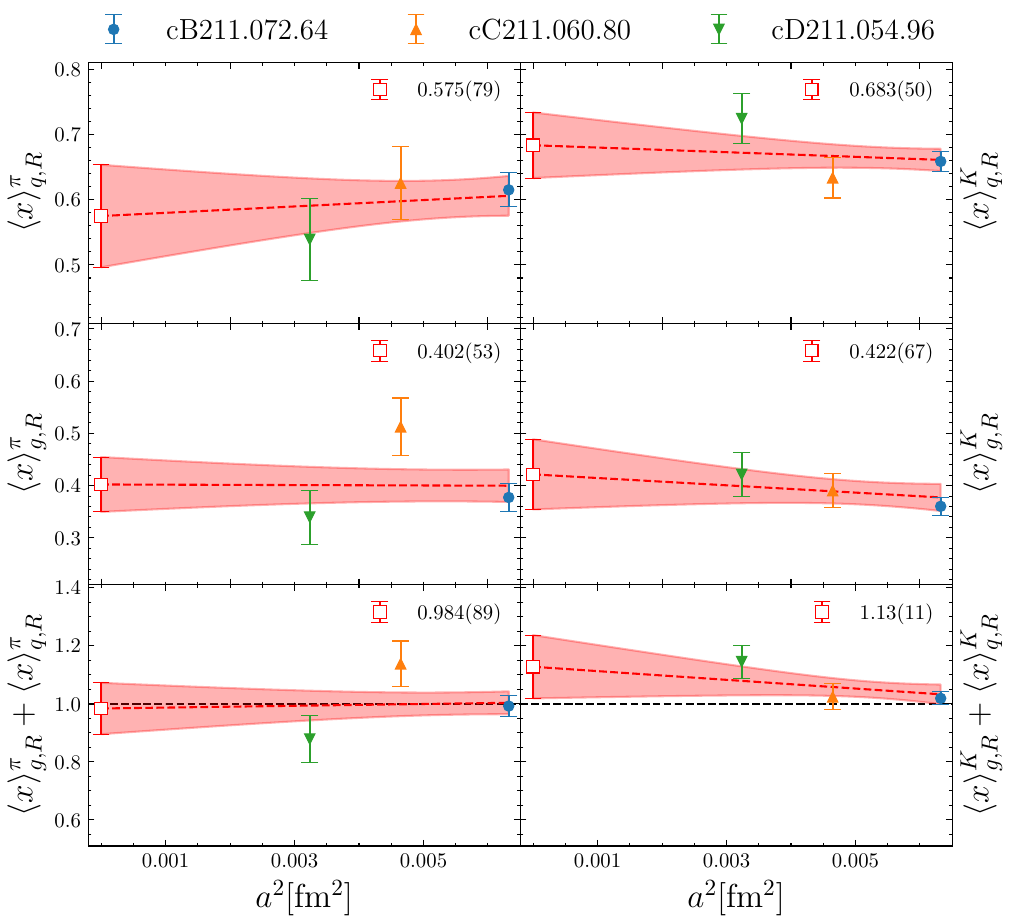}
    \caption{Continuum limit extrapolation for the pion (left panel) and the kaon (right panel). We present our results for the total quark and gluon contributions, as well as the momentum sum rule. The blue filled circles are results for the B ensemble~\cite{ExtendedTwistedMass:2021rdx}, the orange upwards triangles for the C ensemble and the green downwards triangles for the D ensemble. The open symbol is the result after model averaging over the constant and linear fits.}%{\bf Dina and varying the number of $\tins$ included in the fits??}
    \label{fig:continuum_limit}
\end{figure}

\begin{figure}
    \centering
    \includegraphics[width=\linewidth]{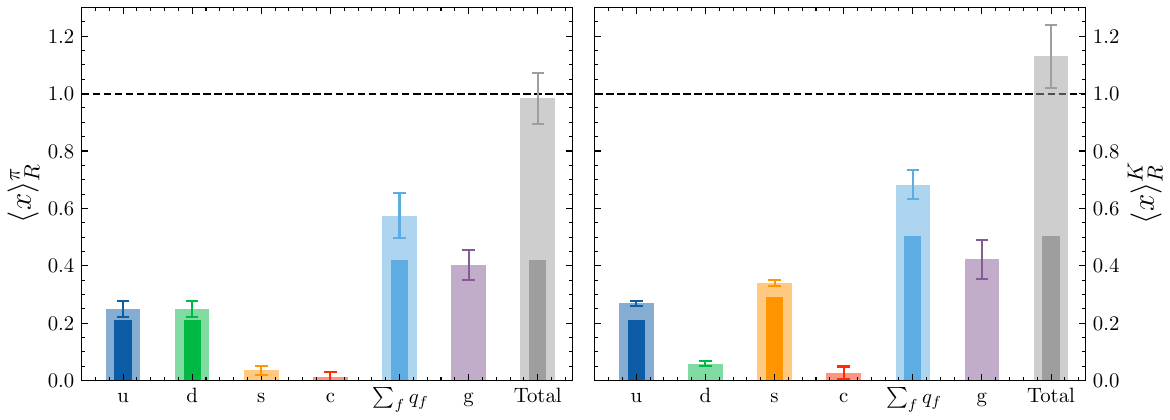}
    \caption{The quark and gluon momentum fractions for the pion (left panel) and kaon (right panel) using the numbers in Table~\ref{table:results}. Inner bars represent only the connected contributions, while the outer bars show the total,  including disconnected contributions.}
    \label{fig:hist}
\end{figure}

The twisted-mass fermion lattice action at maximal twist exhibits automatic $\mathcal{O}(a)$ improvement, meaning that leading discretization artefacts appear at second order in the lattice spacing. In Fig.~\ref{fig:continuum_limit}, we show the results for the three ensembles for the  total quark and  gluon contribution and their sum. The final continuum limit value is obtained by model averaging  the linear fit in $a^2$ and constant fits. The sum rule
\begin{equation}
    \langle x \rangle_{q, R} + \langle x \rangle_{g, R} = 1,
\end{equation}
is satisfied within errors. The results of all renormalized results are given in Table~\ref{table:results} for both the pion and kaon. For comparison, we also include the momentum faction of the nucleon computed using only the B ensemble~\cite{Alexandrou:2020sml}.  As can be seen, the gluon contribution for these three hadrons, is the same within errors. In Fig.~\ref{fig:hist}, we show the momentum fraction for each quark flavor and the gluon as well as  their sum in the continuum limit using the results of Table~\ref{table:results}.

\begin{table}
  \caption{Compilation of results for the pion and the kaon in the continuum limit along those for the proton for the B-
ensemble~\cite{Alexandrou:2020sml}. All quantities are presented at the scale
  $2\ \mathrm{GeV}$ in the $\MSbar$ scheme.}
  \centering
  \begin{tabular*}{0.75\textwidth}{@{\extracolsep{\fill}}lccc}
    \toprule\hline
    & $\pi$ & $K$  & $p$ (B-ensemble)\\
    \midrule\hline
    \(\langle x\rangle_{u,\mathrm{R}}\) & $0.249(28)$ & $0.269(09)$ & $0.354(30)$\\
    \(\langle x\rangle_{d,\mathrm{R}}\) & $0.249(28)$ & $0.059(09)$ & $0.188(19)$ \\
    \(\langle x\rangle_{s,\mathrm{R}}\) & $0.036(15)$ & $0.339(11)$ & $0.052(12)$\\
    \(\langle x\rangle_{c,\mathrm{R}}\) & $0.013(16)$ & $0.028(21)$ & $0.019(09)$ \\
    \(\langle x\rangle_{g,\mathrm{R}}\) & $0.402(53)$ & $0.422(67)$ & $0.427(92)$ \\
    \(\langle x\rangle_{q,\mathrm{R}}\) & $0.575(79)$ & $0.683(50)$ & $0.618(60)$ \\
    \(\langle x \rangle_{u+d-2s,\mathrm{R}}\) & $0.438(18)$ & $-0.362(08)$ & $---$\\
    \(\langle x \rangle_{u+d+s-3c,\mathrm{R}}\) & $0.521(51)$ & $0.494(36)$ & $---$\\
    \(\langle x\rangle_{\mathrm{g, R}} + \langle x\rangle_{\mathrm{q, R}}\) & $0.984(89)$ & $1.13(11)$ & $1.04(11)$ \\    
    \bottomrule\hline
  \end{tabular*}
  \label{table:results}
\end{table}

\begin{figure}
    \centering
    \includegraphics[width=0.6\linewidth]{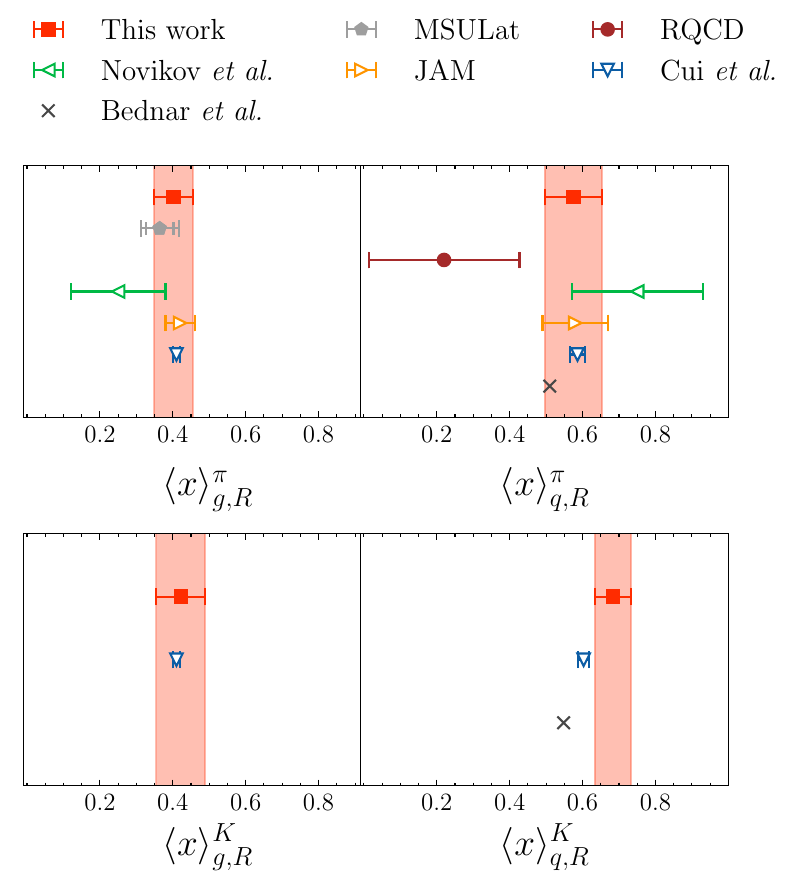}
    \caption{Comparison of the results of this work, with other available data, both from phenomenology and from lattice QCD.  All results are given in the $\MSbar$ scheme at the scale of $\mu=2\ \mathrm{GeV}$. The upper panels show the results for gluon (left) and quark (right) momentum fractions for the pion, $\langle x \rangle_\textrm{g,R}^\pi$ and  $\langle x \rangle_\textrm{q,R}^\pi$, respectively. The lower panels show the corresponding results for the kaon.  The red filled squares show the results of this work with the red band the associated error band.  Recent results from phenomenological analyses of PDFs data are given by open symbols: left green triangle from Ref~\cite{Novikov:2020xft}) and  right orange triangle by the JAM Collaboration~\cite{Barry:2021qpu}. The result based on the LFWF\cite{Cui:2020kaon} is represented by the down blue triangle, while the result from the DSE~\cite{Bednar:2020dist} is represented by the black cross, where no error is provided. Recent lattice QCD results extrapolated to the continuum limit are given by  the brown filled circle (RQCD~\cite{Loffler:2021afv}) and the gray pentagon (MSULat~\cite{Good:2023ecp}).}
    \label{fig:comparison}
\end{figure}

In Fig.~\ref{fig:comparison}, we show a comparison of our results with those of other groups.  We include recent results
from  phenomenological analyses~\cite{Novikov:2020xft,Barry:2021qpu},
from the Dyson Schwinger equations (DSE)~\cite{Bednar:2020dist},
from a calculation using the light-front wave function (LFWF)  approach~\cite{Cui:2020kaon}
and from lattice QCD computations~\cite{Loffler:2021afv,Good:2023ecp},
limiting ourselves to those which are extrapolated to the
 continuum limit.
The two other lattice QCD results shown are computed using
ensembles simulated with pion masses larger than physical
and then extrapolated to the physical pion mass. Furthermore,
they both  considered only a partial sum of disconnected contributions,
namely only the quark disconnected is considered in Ref.~\cite{Loffler:2021afv}
but not the  gluon contribution, whereas only the gluon is considered  in Ref.~\cite{Good:2023ecp}.
There are cases  where all contributions are included, but those are restricted
to only one lattice spacing~\cite{ExtendedTwistedMass:2021rdx}
at physical pion mass or for  larger than physical
pion masses~\cite{Hackett:2023nkr, Alexandrou:2020gxs}.
Our result for $\langle x \rangle_{g,R}^{\pi}$ is in agreement
with the one from Ref.~\cite{Good:2023ecp}, while $\langle x \rangle_{q,R}^{\pi}$ is consistent with RQCD~\cite{Loffler:2021afv}, that, however, carries a very large error.
In the case of the kaon, the only available lattice QCD data~\cite{Alexandrou:2020gxs}
is for one lattice spacing using an ensemble with a heavier-than-physical
pion mass and thus one cannot directly compare to
the present work.

\begin{figure}
    \centering
    \includegraphics[width=\linewidth]{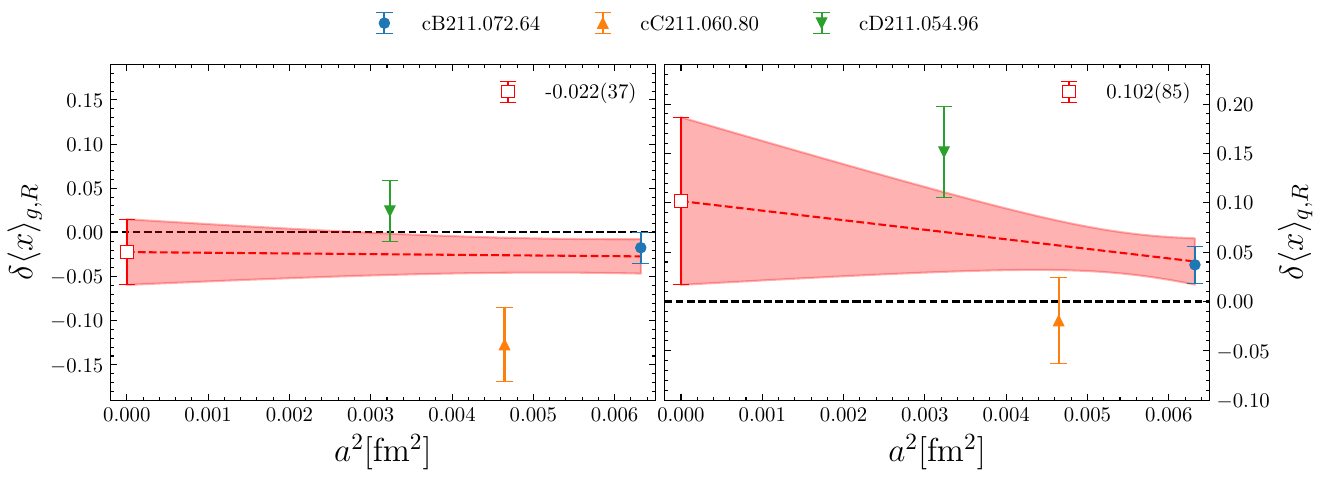}
    \caption{Results on $\delta x_{g,R}$ (left) and $\delta x_{q,R}$ (right).}
    \label{fig:delta}
\end{figure}

For completeness, we calculated the difference between the quark and gluon momentum fractions in the pion and kaon, $\delta x_{q,R}\equiv \langle x\rangle^K_{q,R}-\langle x\rangle^\pi_{q,R}$ and $\delta x_{g,R}\equiv \langle x\rangle^K_{g,R}-\langle x\rangle^\pi_{g,R}$  In Fig.~\ref{fig:delta}, we  show results on 
$\delta x_{q,R}$ and $\delta x_{g,R}$. We  find that $\delta x_{g,R}=-0.022(37)$ and $\delta x_{q,R}=0.102(85)$, which suggest that the gluon  momentum fraction is the same in the  pion and kaon. 

\section{Summary and Conclusions}

We present the flavor decomposition of the momentum fraction for the  pion and kaon in the continuum limit using three ensembles of $N_f = 2 +1+1$ twisted mass fermions at the physical point.  The results presented indicate a similar gluon momentum fraction in the pion and kaon.
There is also agreement with  the gluon momentum fraction found in the nucleon for the B ensemble. Given the mild dependence on the lattice spacing, one may deduce that the gluon momentum  in proton in the continuum limit will remain similar as that in the pion and kaon.

In the future, we aim to  reduce the statistical errors  both by increasing the number of source positions and gauge configurations. In addition, the analysis of an ensemble closer to the continuum limit will help to better control the continuum extrapolation.

\vspace*{-0.3cm}

\begin{acknowledgments}\vspace*{-0.3cm}
    G.S. acknowledges financial support from the European Regional Development Fund and the Republic of Cyprus through the Cyprus Research and Innovation Foundation under contract number EXCELLENCE/0421/0195.
    This project is partly funded by the European Union’s Horizon 2020 Research and Innovation Programme ENGAGE under the Marie Sklodowska-Curie COFUND scheme with grant agreement No. 101034267. C.A. acknowledges partial support from the  project 3D-nucleon funded by the European Regional Development Fund and the Republic of Cyprus through the Cyprus Research and Innovation Foundation under contract number EXCELLENCE/0421/0043 and the European Joint Doctorate project AQTIVATE funded by the European Commission with the Grant Agreement No 101072344. This work is supported by the Swiss National Science Foundation (SNSF) through grant No.~200021\_175761, 200021\_208222, and 200020\_200424,
    as well as by the DFG and the NFSC as part of the Sino-German Collaborative Research Center CRC110 {\it Symmetries and the emergence of structure}. J. D. and M.~C. acknowledge financial support from the U.S. Department of Energy, Office of Nuclear Physics under Grant No.\ DE-SC0020405, and the Grant No.\ DE-SC0025218.  J.F. is
    supported by the DFG research unit FOR5269 ”Future methods for studying confined gluons
    in QCD".
    We acknowledge computing time granted on Piz Daint at Centro Svizzero di Calcolo Scientifico (CSCS)
    via the projects s849, s982, s1045, s1133 and s1197 and
    the Gauss Centre for Supercomputing e.V. (www.gauss-centre.eu) for providing computing time on the GCS Supercomputers SuperMUC-NG
    at Leibniz Supercomputing Centre and JUWELS \cite{JUWELS} at Juelich Supercomputing
    Centre. The authors acknowledge the Texas Advanced Computing Center (TACC) at
    the University of Texas at Austin for  HPC resources (Project ID PHY21001).
    %%%
    We acknowledge PRACE for awarding access to HAWK at HLRS, project with Id Acid 4886, and the Swiss National Supercomputing Centre (CSCS) and the EuroHPC Joint Undertaking for awarding access to the LUMI supercomputer,through the Chronos programme under project IDs CH17-CSCS-CYP and CH21-CSCS-UNIBE as well as the EuroHPC Regular Access Mode under project ID EHPC-REG-2021R0095.
    The open source software packages
    tmLQCD~\cite{Jansen:2009xp,Abdel-Rehim:2013wba,Deuzeman:2013xaa},
    Lemon~\cite{Deuzeman:2011wz},
    QUDA~\cite{Clark:2009wm,Babich:2011np,Clark:2016rdz}, R~\cite{R:2019},
    cvc \cite{CVC:2024} and plegma have been used. Finally,
    We thank the authors of~\cite{Cui:2020kaon} and~\cite{Bednar:2020dist} for kindly sending us
    their results evolved to the scale used in our
    computation, $2\ \mathrm{GeV}$.

\end{acknowledgments}

\end{document}